\documentclass[12pt,preprint]{aastex}
\slugcomment{{\sc Accepted to ApJS:} July 22, 2013} 
\usepackage{natbib}
\usepackage{amsmath}

\shorttitle{The Continuum Linear Polarization in Classical Be Stars}
\shortauthors{Halonen \& Jones}

\begin{document}

\title{On the Intrinsic Continuum Linear Polarization of Classical Be Stars: The Effects of Metallicity and One-Armed Density Perturbations}		
\author{Robbie J. Halonen \& Carol E. Jones}
\affil{Department of Physics and Astronomy, The University of Western Ontario, London, ON, Canada, N6A 3K7}
\email{rhalonen@uwo.ca}
\accepted{23/07/13}

\begin{abstract}
We report on the effects of two disk properties on the continuum linear polarization signature of classical Be stars. First, we investigate the effect of including metallicity in computing the thermal structure of the circumstellar gas on the resulting polarimetric Balmer jump. The Balmer jump is a distinguishing feature of the polarization signature in these objects and, as such, can be used as a tool for differentiating classical Be stars from similar H$\alpha$-emitters identified through conventional photometric techniques. We find that although low-metallicity environments will have hotter disk temperatures on average, the temperature change alone cannot account for the discrepancy in the frequency of Balmer jumps between low-metallicity and solar-metallicity stellar populations. Second, we investigate the effect of including a global one-armed oscillation in the gas density distribution of the modeled disk. We find that a non-axisymmetric perturbation pattern yields discernible variations in the predicted polarization level. If these density oscillations are present in the inner region of classical Be star disks, the polarimetric variations should produce a periodic signature which can help characterize the dynamical nature of the gas near the star.
\end{abstract}

\keywords{Methods: numerical --- Polarization --- Scattering --- (Stars:) circumstellar matter --- Stars: emission-line, Be}

\section{Introduction}

Classical Be stars are objects that exhibit, or have exhibited at some point in time, observational properties indicative of a Keplerian disk of ionized gas orbiting a non-supergiant star. These objects are characterized by features that arise from the interaction between the radiation emitted by the massive central star and the enveloping material, most prominent among which are an emission-line spectrum, an excess of continuum emission, and a linear polarization signature. While the steady-state viscous disk model \citep{lee91,oka01} provides a good theoretical description of the geometrically-thin, Keplerian disk from which these features originate, the mechanisms through which the viscous disk is fed gas and angular momentum remain unidentified. Also unclear are the reasons for which particular B stars undergo the classical Be phenomenon while others do not. Almost certainly, the rapid rotation of these objects plays an important role in their development, but the accuracy of rotational velocity determinations is contentious \citep{tow04} and debate regarding the extent to which rotation can drive the formation of the decretion disks persists. Identifying the processes which may be responsible for producing tenable Keplerian decretion disks around rapidly-rotating B-type stars remains an intriguing challenge \citep[for further discussion, see the review papers by][]{por03,owo06}.

Determining the origins of the classical Be phenomenon requires a reliable understanding of the physical and dynamical nature of the gaseous envelopes. While our current understanding is unsatisfactory in many regards, steady progress is being achieved through the detailed modeling of high-resolution observations using modern non-LTE radiative transfer codes, such as those described by \citet{car06} and \citet{sig07}. Beginning with our work in \citet{hal13b}, we have considered the modeling of the polarimetric properties of classical Be stars for tracing the evolution of the circumstellar gas. In this report, we supplement our previous investigation by considering two important disk properties which may have an appreciable effect on the computation of polarimetric observables. First, we investigate the effect of gas metallicity on the thermal solution of the modeled disk. As the gas temperature can profoundly affect the state of the scattering and absorptive opacities in the disk, it is essential to account for effects which may cause discernible changes to the computed solution. Second, we consider the implications of including non-axisymmetric density perturbations in the disk. Such perturbations alter the physical and geometric properties of the region in which electron scattering, the process responsible for the linear polarization signature in classical Be stars, occurs. For both the inclusion of gas metallicity and the addition of density perturbations, we determine how these modifications to the model affect the properties of the disk and, as a result, the predictions of the continuum linear polarization.

In this paper, we evaluate the importance of gas metallicity and non-axisymmetric density distributions in the computation of the continuum linear polarization signature from classical Be star models. While these two disk properties are unrelated in most respects, understanding how each property affects the circumstellar gas is important when modeling and interpreting observables. For this reason, they are both key considerations in our ongoing analysis of the properties of the linearly polarized light observed in classical Be stars. We use the Monte Carlo calculation of the Stokes intensities described in \citet{hal13a} with the self-consistent thermal solution of the non-LTE radiative transfer code of \citet{sig07} to predict the polarimetric observables. We have organized the paper as follows: in Section 2, we briefly explain the computational procedures used in this investigation; in Section 3, we compare the polarimetric Balmer jump from classical Be stars with low- and solar-metallicity compositions; in Section 4, we consider the presence of one-armed density oscillations in the disk; lastly, in Section 5, we summarize the findings of this report.

\section{Computational Method}

We obtain the linear polarization for classical Be star models computed using the non-LTE radiative transfer code developed by \citet{sig07} and the Monte Carlo multiple-scattering treatment described in \citet{hal13a}. The Sigut \& Jones code solves the coupled problems of statistical and radiative equilibrium to produce a self-consistent calculation of the thermal structure of the circumstellar disk. The Monte Carlo procedure adopts the computed atomic level populations and gas temperatures as the underlying model of the circumstellar envelope and simulates the propagation of photons from the photoionizing radiation field of a star described by a \citet{kur93} model atmosphere. Random sampling is used for determining the photon path lengths and scattering angles, and for resolving photon interactions with the gas. Combining the radiative equilibrium solution of the Sigut \& Jones code with the Monte Carlo simulation provides an effective computational procedure for modeling circumstellar disks, such as those of classical Be stars, while employing realistic chemical compositions and self-consistent calculations of the thermal structure of the disk.

The density distribution of the circumstellar gas follows a simple power-law parameterization first prescribed to explain IR observations of Be stars \citep{wat86,wat87}. The density of the gas at coordinates $(R, Z)$, where $R$ is the radial distance from the stellar rotation axis and $Z$ is the distance above the midplane, is specified by 
\begin{equation} \label{eq:dens}
 \rho(R,Z) = \rho_{0}(R)^{-n}e^{-(\frac{Z}{H})^2}.
\end{equation} 
Here, $\rho_0$ fixes the density of the disk at the surface of the central star, $n$ sets the exponential decline in density with increasing distance from the star, and $H$ establishes the vertical scale height at each radial point using an initial value for the gas temperature, typically around 0.6 $T_{\textrm{eff}}$. In this work, our disk models use 60 radial grid points extending from the stellar surface outward to 100 stellar radii. At each $R$, the gas is assumed to be in vertical hydrostatic equilibrium perpendicular to the plane of the disk. The models use 30 vertical grid points above and below the midplane. The Monte Carlo simulation extends the two-dimensional computational domain from the Sigut \& Jones code into a three-dimensional model using 72 azimuthal grid points. While observational investigations have found that the value of $n$ can range from 2.0 to 5.0, we use $n=3.5$ which is the value predicted from isothermal viscosity \citep{por99}. For the purposes of this work, we considered main sequence stars of spectral types B0 through B8. The stellar parameters used for each model are listed in Table \ref{tab:1}.

\begin{table}
\begin{center}
\caption{Stellar Parameters}
\label{tab:1}
\begin{tabular}{lccccc}
 \tableline \tableline
Spectral & Radius & Mass & Luminosity & $T_{\rm eff}$ & $\log(g)$\\
Type &($R_{\sun}$)&($M_{\sun}$)&($L_{\sun}$)&(K)& ($\rm cm\, s^{-2}$)\\
\tableline
B0V & 7.40 & 17.5 &  3.98 $\times 10^{4}$ & 3.00 $\times 10^{-4}$ & 3.9\\
B1V & 6.42 & 13.2 &  1.45 $\times 10^{4}$ & 2.54 $\times 10^{-4}$ & 3.9\\
B2V & 5.33 & 9.11 &  4.76 $\times 10^{3}$ & 2.08 $\times 10^{-4}$ & 3.9\\
B3V & 4.80 & 7.60 &  2.58 $\times 10^{3}$ & 1.88 $\times 10^{-4}$ & 4.0\\
B4V & 4.32 & 6.62 &  1.33 $\times 10^{3}$ & 1.68 $\times 10^{-4}$ & 4.0\\
B5V & 3.90 & 5.90 &  7.28 $\times 10^{2}$ & 1.52 $\times 10^{-4}$ & 4.0\\
B6V & 3.56 & 5.17 &  4.31 $\times 10^{2}$ & 1.38 $\times 10^{-4}$ & 4.0\\
B7V & 3.28 & 4.45 &  2.28 $\times 10^{2}$ & 1.24 $\times 10^{-4}$ & 4.1\\
B8V & 3.00 & 3.80 &  1.36 $\times 10^{2}$ & 1.14 $\times 10^{-4}$ & 4.1\\
\tableline
\end{tabular}
\tablecomments{Based on values from \citet{cox00}.}
\end{center}
\end{table}

The source of the intrinsic continuum polarization in classical Be stars is Thomson scattering \citep[for example][]{coy69,zel72}. When unpolarized light is scattered by free electrons, it becomes linearly polarized perpendicular to the plane containing the incident and scattered radiation. If the scattering source is spherically symmetric on the plane of the sky, the distribution of polarizing planes is uniform, a complete cancellation of vibrations from orthogonal directions occurs and the net polarization is zero. Thus, axisymmetric disks, as an example, should exhibit zero net polarization when viewed pole-on. The projection of the disk on the plane of the sky is spherically asymmetric at any other viewing angle. The polarization level arising from electron scattering in the disk increases with inclination, peaking at around $75^{\circ}$ \citep{hal13b}. At inclinations higher than $75^{\circ}$, the polarization level declines as the scattered light undergoes increased attenuation from H{\sc i} absorption. 

The Monte Carlo procedure determines the fraction of polarized light emerging from the modeled circumstellar environment by calculating the Stokes intensities, which are simply the sums or differences in intensity between components of the radiation field measured along the defining axes. The $I$ parameter is the total intensity of the beam of light. The $Q$ and $U$ parameters, representing the linear polarization, are the differences in intensity between orthogonal directions of vibration, with one pair of orthogonal axes fixed at 45$^{\circ}$ from the other. As electron scattering produces no circularly polarized component, we ignore the $V$ parameter and the normalized polarization level is expressed as:
\begin{equation}
p = (q^2+u^2)^{1/2}
\end{equation}
where $q = Q/I$ and $u = U/I$. In an axisymmetric disk, one of the linear polarization parameters is always zero if one defines the principal axes to coincide with the plane of the disk when viewed at $90^{\circ}$. Accounting for both linear polarization parameters becomes necessary when the density distribution in the disk is not axially symmetric.

\section{Metallicity}

The origin of the classical Be phenomenon may be intrinsically tied to the evolutionary characteristics of the stars that constitute this group.  As such, the respective roles of metallicity and stellar age are important focuses for classical Be star research. Comparison studies of the fractional Be star populations of Milky Way (MW), Small Magellanic Cloud (SMC) and Large Magellanic Cloud (LMC) clusters suggest that the prevalence of the Be phenomenon increases in lower metallicity environments \citep{mae99}. This result carries important implications for the nature of classical Be stars as metallicity may have a dominant effect on the fraction of stars that attain critical rotation \citep{mae01}. The evolutionary age of Be stars remains somewhat unsettled due to the conflicting suggestions that (1) the phenomenon occurs largely in the latter half of the main sequence \citep{fab00} and (2) the phenomenon is present throughout the entire main sequence \citep{mat08,wis06}. Presenting perhaps the most compelling research on the issue to date, \citet{wis07b}'s investigation Be stars in the LMC and SMC suggests that classical Be stars are present in young stellar clusters with an enhanced fraction in older clusters. Of significance, the authors point out that two-colour photometry, the method commonly used to study evolutionary age and metallicity in the Be phenomenon, cannot reliably discern between classical Be stars and other B-type H$\alpha$ emitters. They refine the classical Be identification process through the use of these objects' distinct linear polarization signature characterized by the wavelength-dependent imprint of neutral hydrogen absorption.

In this section, we evaluate the importance of metallicity in computing the thermal structure of Be star disks and the effect of metallicity-induced temperature differences on the predicted polarimetric Balmer jump. In light of \citet{wis07b}'s finding that the frequency of polarimetric Balmer jumps is smaller in low metallicity environments, we address the possibility that this is due to the inherent temperature discrepancies between disks in different metallicity environments, as reported by \citet{ahm11}.

\subsection{Results and Discussion}

\begin{figure*}
\epsscale{1.0}
\plotone{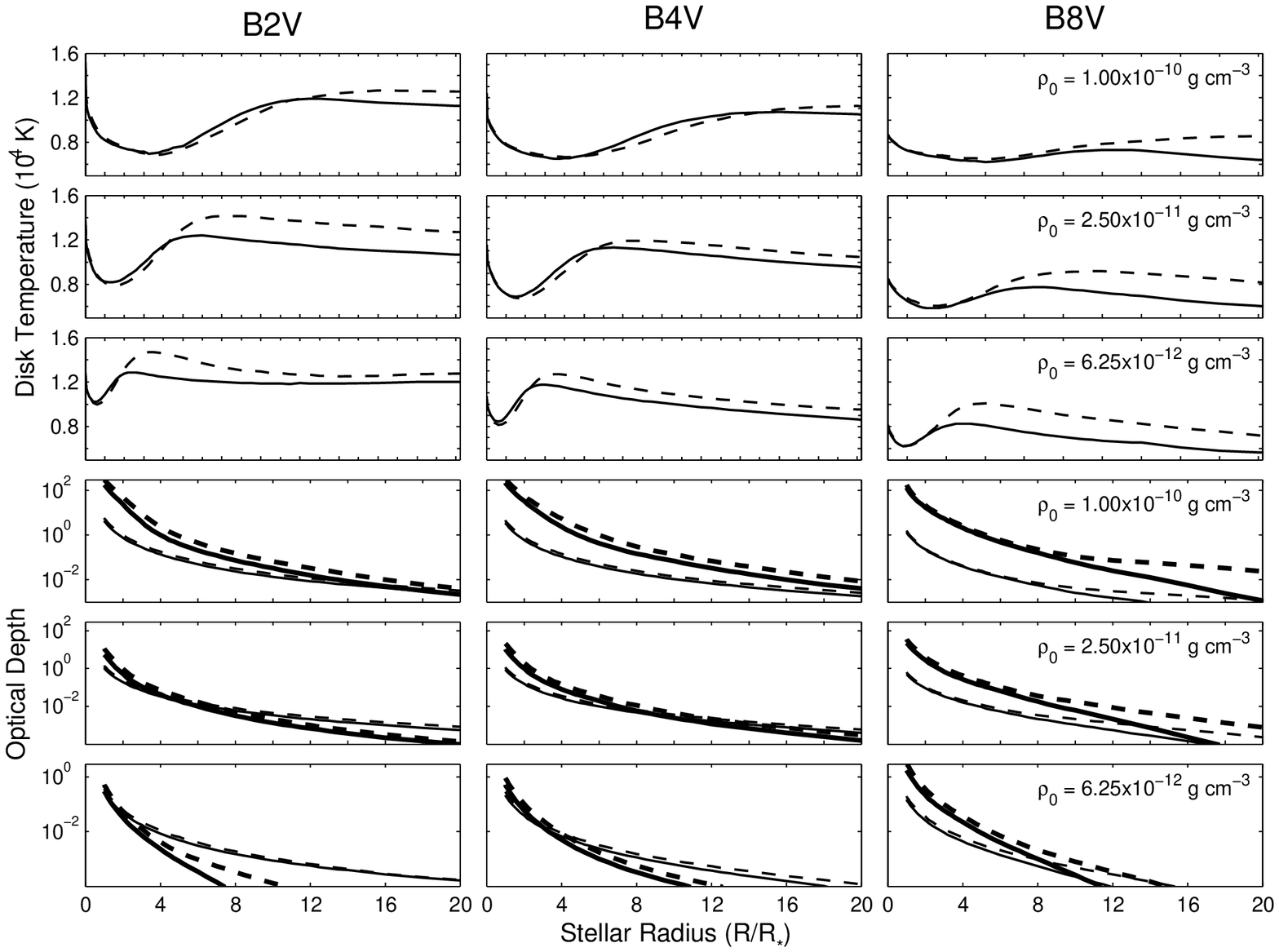}
\caption{Disk temperature and optical depth comparison of circumstellar disks with $n = 3.5$ and $\rho_0 = 1.0 \times 10^{-10}$ g cm$^{-3}$ (top frame), $\rho_0 = 2.5 \times 10^{-11}$ g cm$^{-3}$ (middle frame), $\rho_0 = 6.25 \times 10^{-12}$ g cm$^{-3}$ (bottom frame). The models on the left are disks surrounding a B2V star, those in the center are disks surrounding a B4V star, and those on the right are disks surrounding a B8V star. The dashed lines represent a pure hydrogen composition and the solid lines represent a solar chemical composition. In the bottom frames, the thick lines show the optical depth from hydrogen absorption just shortward of the Balmer jump, while the thin lines show the optical depth from electron scattering.}
\label{fig:1}
\vspace{0.1in}
\end{figure*}

We began by computing two sets of models using identical mass density distributions and different gas compositions. One set of models used a pure hydrogen composition while the second set included nine elements (H, He, C, N, O, Mg, Si, Ca and Fe) and assumed solar abundances. The details of the atomic models implemented in the radiative equilibrium calculation and the exact chemical abundances adopted are given in \citet{sig07}. The comparison of these two sets of models provide us with a reasonable gauge to the effect of including heating and cooling from metal processes in the determination of the thermal structure of the disk. We evaluate the global disk temperature using a density-weighted temperature defined as:
\begin{equation}
\langle T_\rho \rangle =\dfrac{1}{M_{disk}} \int T(R,Z) \rho(R,Z) dV ,
\end{equation}
where M$_{disk}$ is the total mass of the disk. We find that the density-weighted temperature does not differ appreciably between the two sets, as shown in Table \ref{tab:2}. This is consistent with the findings of \citet{jon04} who reported very small differences between the self-consistent density-weighted temperatures of pure hydrogen and solar Fe abundance disks for models of the classical Be stars $\gamma$ Cas and 1 Del. \citet{sig07} noted that the absorption of ionizing radiation by elements heavier than hydrogen offsets some of the line cooling, arising mainly from Fe\textsc{ii}, in higher density disks.  Figure \ref{fig:1} shows the density-weighted temperature of the gas as a function of radial distance for three stars. We note that the gas temperatures in all the sets are very similar within the first six stellar radii of the disk, the region of the disk where the polarimetric Balmer jump forms \citep{hal13b}. While the temperatures differ somewhat more in the inner part of the disk for the models where $\rho_0 = 6.25 \times 10^{-12}$ g cm$^{-3}$, the density of the disk in this model yields insufficient absorption for producing a Balmer jump in the polarization spectrum. From these temperature comparisons, we do not expect there to be significant differences in the polarimetric properties caused by the inclusion of metallicity in the radiative equilibrium computation.

\begin{table*}
\begin{center}
\caption{Density-Weighted Disk Temperatures}
\label{tab:2}
\begin{tabular}{lccc}
 \tableline \tableline
 & & & \\
$\rho_0$ & $6.25 \times 10^{-12}$ g cm$^{-3}$ & $2.50 \times 10^{-11}$ g cm$^{-3}$ & $1.00 \times 10^{-10}$ g cm$^{-3}$\\
& & & \\
\hline \hline
B2V &  &  &  \\
Hydrogen & 1.22 $\times 10^{4}$ K & 1.09 $\times 10^{4}$ K & 9.49 $\times 10^{3}$ K  \\
Solar & 1.16 $\times 10^{4}$ K & 1.04 $\times 10^{4}$ K & 9.29 $\times 10^{3}$ K  \\ \hline \hline
B4V &  &  &  \\
Hydrogen & 9.51 $\times 10^{3}$ K & 9.00 $\times 10^{3}$ K & 8.61 $\times 10^{3}$ K    \\
Solar & 9.00 $\times 10^{3}$ K & 8.60 $\times 10^{3}$ K & 8.28 $\times 10^{3}$ K   \\ \hline \hline
B8V &  &  &  \\
Hydrogen & 7.47 $\times 10^{3}$ K & 7.25 $\times 10^{3}$ K & 7.20 $\times 10^{3}$ K   \\
Solar & 6.84 $\times 10^{3}$ K & 6.68 $\times 10^{3}$ K & 6.60 $\times 10^{3}$ K    \\ \hline \hline
\tableline
\end{tabular}
\end{center}
\end{table*}

\begin{figure}
\epsscale{1.0}
\plotone{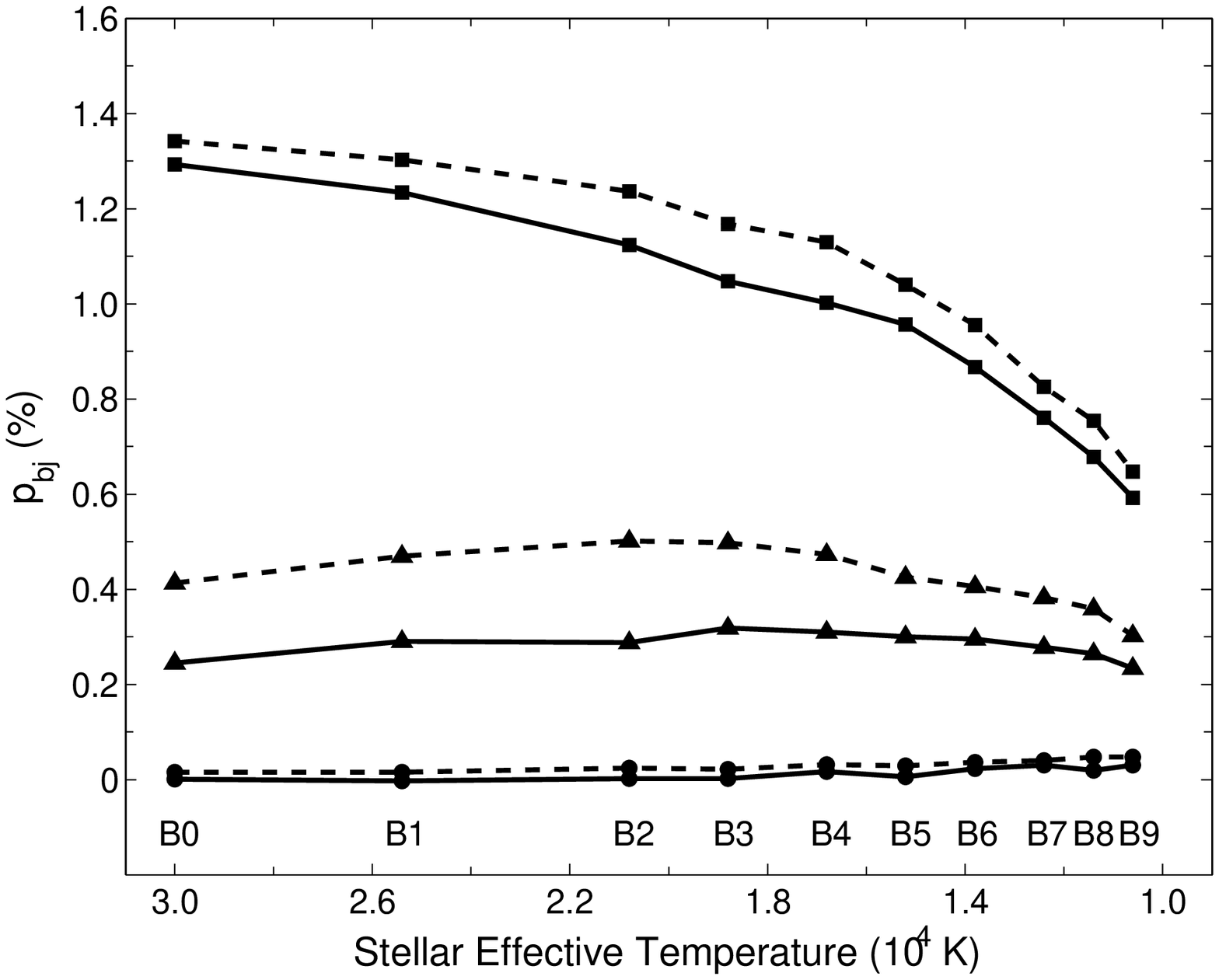}
\caption{Polarimetric Balmer jump for stars surrounded by circumstellar disks of gas while keeping mass density fixed in models using a pure hydrogen composition (dashed lines) and solar composition (solid lines). The disk density distributions have $n = 3.5$ and $\rho_0 = 1.0 \times 10^{-10}$ g cm$^{-3}$ (squares), $\rho_0 = 2.5 \times 10^{-11}$ g cm$^{-3}$ (triangles), and $\rho_0 = 6.25 \times 10^{-12}$ g cm$^{-3}$ (circles). The system is viewed at an inclination of $i$ = $75^{\circ}$. }
\label{fig:2}
\vspace{0.1in}
\end{figure}

A consequence of the fixed-density scheme is that the hydrogen populations differ between the compared sets of models. As such, the relevant opacities that govern the linear polarization signature, the electron scattering opacity and the H{\sc i} absorptive opacity, are affected by this difference when computing the underlying models. The general result is that the opacities are lower in the solar metallicity model, as demonstrated in Figure \ref{fig:1}. Plainly, this effect arises from the choice of using the same mass density for both sets of models: the solar metallicity model includes less hydrogen and has fewer free electrons available for scattering. With this important consideration in mind, we examine how the Balmer jump changes due to this simple but important difference in the models. We compute the Balmer Jump by taking the difference in the continuum polarization in 100\AA~ bands on either side of the limit, centred roughly at 3600\AA~ and 3700\AA~ respectively. Figure \ref{fig:2} shows the polarization Balmer jump for both sets. The higher opacities in the hydrogen-only models contribute to systemically higher Balmer jumps. Clearly, caution must be taken in the modeling of polarimetric features: a hydrogen-only composition in the disk, as is often employed, can overestimate the amount of electron scattering and H{\sc i} absorption in the disk.

\begin{figure}
\epsscale{1.0}
\plotone{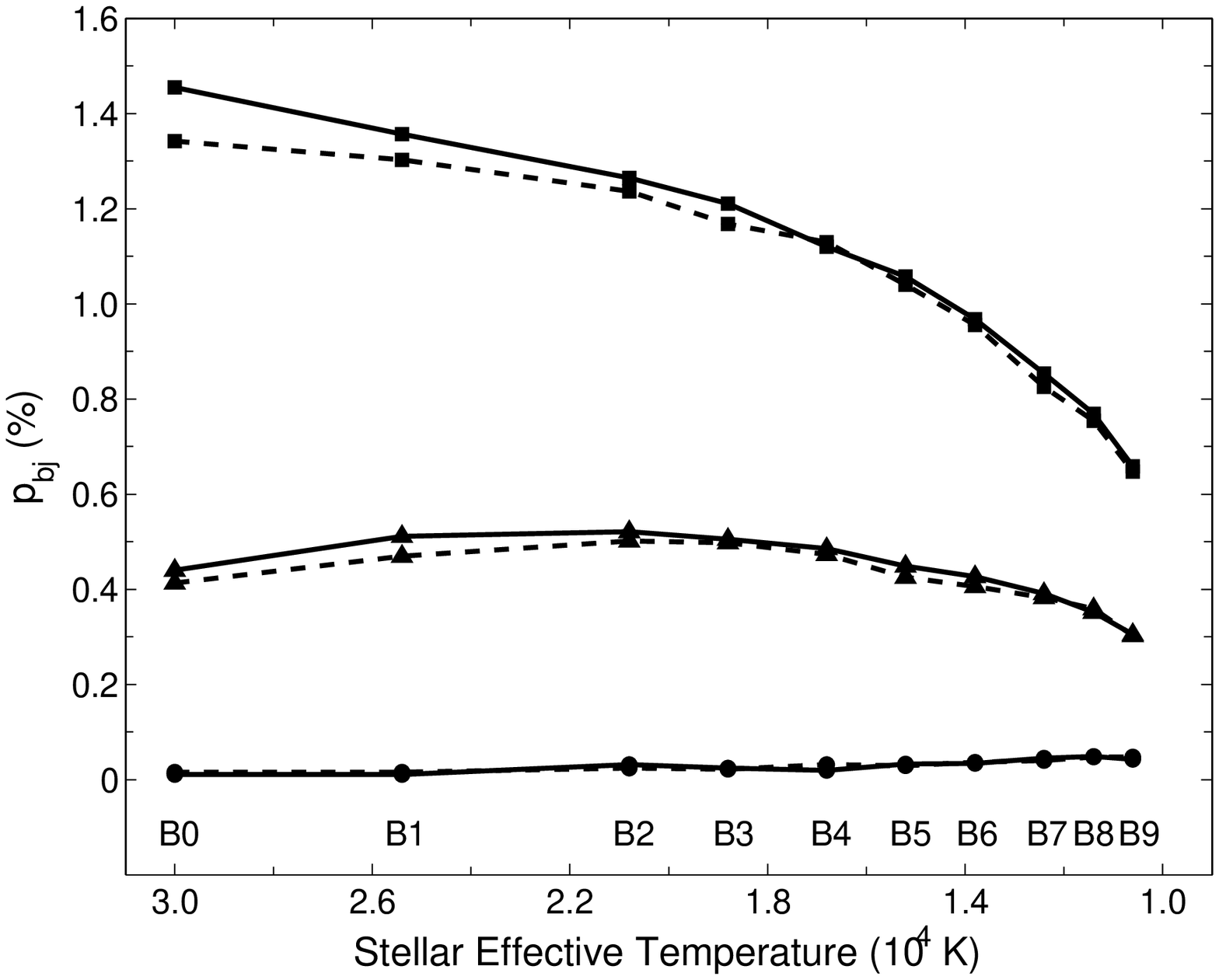}
\caption{Polarimetric Balmer jump for stars surrounded by circumstellar disks of gas while keeping hydrogen number density fixed in models using a pure hydrogen composition (dashed lines) and solar composition (solid lines). The disk density distributions have $n = 3.5$ and $\rho_0 = 1.0 \times 10^{-10}$ g cm$^{-3}$ (squares), $\rho_0 = 2.5 \times 10^{-11}$ g cm$^{-3}$ (triangles), and $\rho_0 = 6.25 \times 10^{-12}$ g cm$^{-3}$ (circles). The mass densities of the solar composition models are adjusted to the fixed hydrogen number density. The system is viewed at an inclination of $75^{\circ}$.}
\label{fig:3}
\vspace{0.1in}
\end{figure}

In another set of models, we fixed the hydrogen number density and performed the same comparison as above. Again, the global disk temperature between the solar-metallicity and pure hydrogen models did not differ significantly close to the star and hence there were no appreciable differences in the scattering and hydrogen absorptive opacities. The Balmer jumps between the two models were nearly identical as illustrated in Figure \ref{fig:3}. Given sufficient density for the characteristic hydrogen absorption signature to be discernible ($\rho_0 \gtrsim 1.0 \times 10^{-11}$ g cm$^{-3}$ for $n = 3.5$), the polarimetric Balmer jump reflects primarily the amount of neutral hydrogen absorption occurring in the disk. As we have shown, in densities above this threshold, the presence of metal-line cooling in the disk does not sufficiently affect the thermal structure to produce an observable change in the Balmer jump.

\begin{figure}
\epsscale{1.0}
\plotone{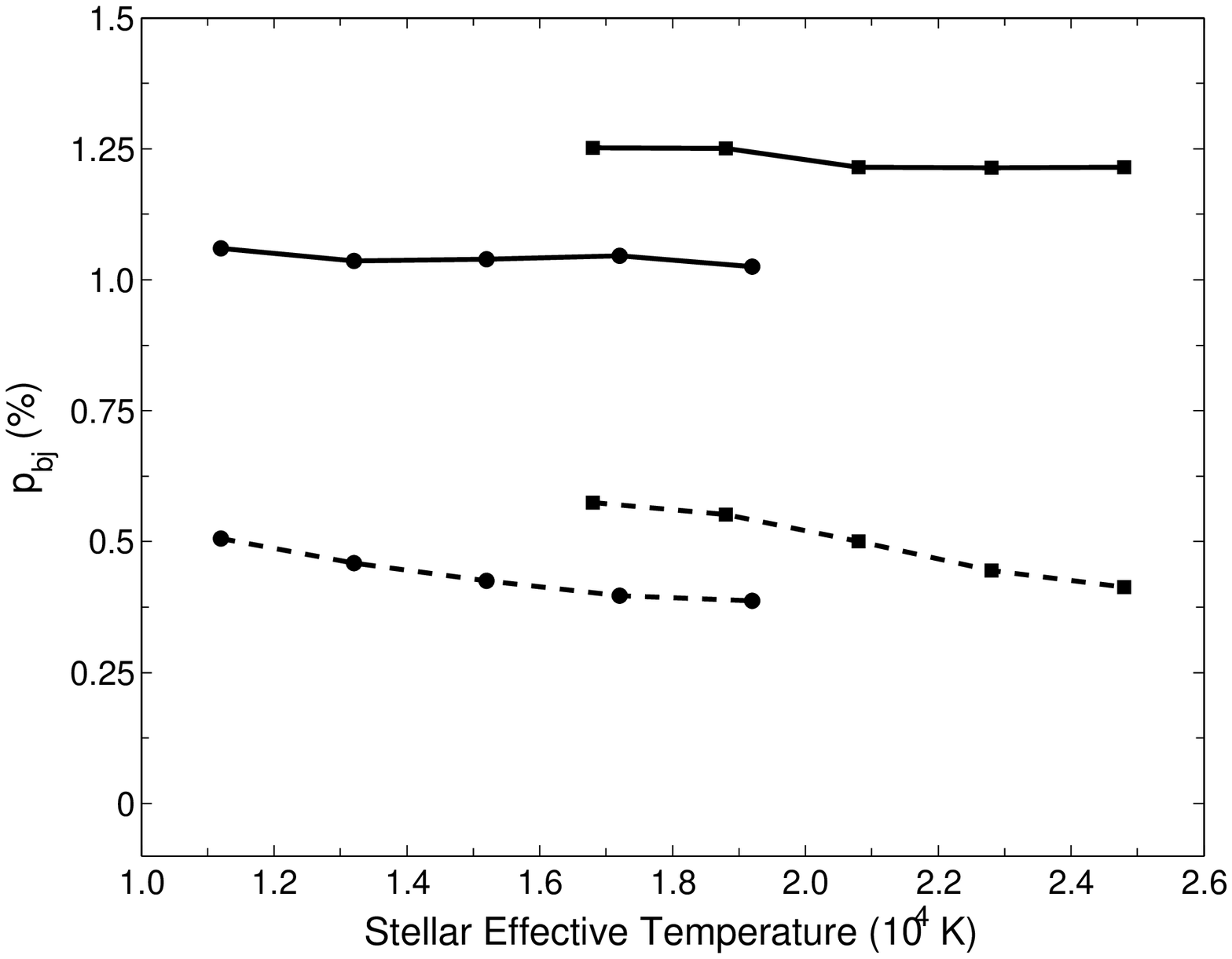}
\caption{Polarimetric Balmer jump for stars using the B2 (squares) and B5 (circles) stellar parameters and varying $T_{\textrm{eff}}$. The disk density distributions are parametrized by $n = 3.5$ and $\rho_0 = 2.5 \times 10^{-11}$ g cm$^{-3}$ (dashed lines) or $\rho_0 = 1.0 \times 10^{-10}$ g cm$^{-3}$ (solid lines) and the systems are viewed at an inclination of $75^{\circ}$.}
\label{fig:4}
\vspace{0.1in}
\end{figure}

Of course, other considerations may produce temperature differences between disks in environments with different metallicities. Using the same non-LTE radiative transfer code to solve for the radiative equilibrium solution as is used in this work, \citet{ahm11} showed that disks in the SMC (\textit{Z} = 0.002) are systematically hotter than disks in the MW (\textit{Z} = 0.02). In their work, the authors accounted for the intrinsic difference in the effective temperatures of spectral types between MW and SMC populations \citep{tru07}. As such, they adopted $T_{\textrm{eff}}$'s for SMC stars that were typically a few thousand degrees higher than the $T_{\textrm{eff}}$'s used for MW stars. In the context of this investigation, we examined the extent to which the intrinsic difference in the $T_{\textrm{eff}}$ could affect the polarimetric Balmer jump.  Figure \ref{fig:4} shows the Balmer jumps for B2V and B5V stars with varying stellar $T_{\textrm{eff}}$ in models computed using the same disk density distribution. In general, the height of the polarimetric Balmer jump decreases with increasing $T_{\textrm{eff}}$ as the gas temperature in the disk increases and the absorptive opacity decreases. However, the changes in the Balmer jump are relatively small. For the optically thinner model, the largest decrease in the Balmer jump over a 2000 K increase in $T_{\textrm{eff}}$ is roughly 10 \%. In the optically thicker model, there was almost no change in the height of the Balmer jump over the same increase in $T_{\textrm{eff}}$. While the circumstellar disks around stars in low-metallicity environments are hotter, the resulting decrease in the H{\sc i} absorptive opacity does not seem to be sufficient to cause an appreciable difference in the computed Balmer jump.

\section{Non-axisymmetric Disk Density Perturbations}

The outward flow of gas and angular momentum in a Keplerian disk is well-described by viscous transport \citep{lee91,pap92,oka01}. This theoretical description of the gas dynamics adheres well to the observed kinematic properties of the disks of classical Be stars \citep{car11}. Additionally, the viscous disk model accommodates the existence of non-axisymmetric distributions of gas arising in the form of one-armed oscillations \citep{oka91}. These density waves qualitatively explain a frequently observed characteristic of classical Be stars: non-equal intensities in the components of double-peaked emission lines \citep{han95,oka97}. These asymmetries in line profile morphologies are commonly referred to as the V/R (violet-to-red) variability. The perturbation pattern in the disk comprises overdense and underdense regions of gas which produce emission excesses and deficits at wavelengths shifted by the motion of the gas with respect to the observer. As the wave precesses about the central star, the overdense and underdense regions oscillate from approaching to receding and the excess and deficit in line emission fluctuate about center of the line, producing cyclical V/R variations. The one-armed oscillation paradigm is consistent with observed timescales of periodic V/R variability in classical Be stars, which exist on timescales of a couple of years to ten years \citep{oka97}. Interferometric observations \citep{vak98,ber99} have yielded evidence of one-armed oscillations precessing in the prograde direction in the several classical Be stars.

One peculiarity is the occasional discrepancy in the V/R ratio of different emission lines observed contemporaneously. While most V/R asymmetric lines in a spectrum possess the same orientation, there have been reports of phase differences \citep{sle82,baa85} and even complete V/R reversals \citep{cla00}. \citet{wis07a} suggested that this phenomenon can be explained by the presence of a one-armed oscillation possessing a spiral-like shape and the knowledge that the physical sizes of line forming regions vary for different lines. Therefore, attributing some degree of helicity to the perturbation pattern,  line forming regions of different radial extents may possess very different azimuthal morphologies. In a detailed modeling of the classical Be star $\zeta$ Tau, \citet{car09} employed a global one-armed oscillation model with a spiral perturbation pattern to successfully reproduce the phase differences in V/R variable H$\alpha$ and Br$\gamma$ lines. Their result lends support to the idea that the one-armed oscillation pattern in the disks of classical Be stars may take on a spiral structure.

In this section, we evaluate the effect of global one-armed oscillations on the continuous linear polarization spectrum. As the distribution of the scattering material relative to the observer is fundamental to the measured polarization, properly accounting for non-axisymmetric distributions of gas may be important for attempting to derive physical properties from polarimetric observations of classical Be stars. Recall that the contribution to the polarization level from points that are separated azimuthally by 180 degrees is equivalent. Hence, in the single-scattering limit where no absorption occurs, the polarization level remains unaffected by antisymmetric changes to the density distribution. However, when further effects are taken into consideration, such as stellar occultation, multiple-scattering and absorption/re-emission, the result will be variations in the polarization level.

\subsection{Results and Discussion}

We computed models that include global one-armed oscillations in the gas density distribution of the circumstellar disk. These models were constructed by applying a perturbation to the density distribution given by Equation \ref{eq:dens}. The thermal solution was calculated using the Sigut \& Jones code for 36 azimuthal angles in equal intervals from $0$ to $2\pi$. The computed atomic level populations and gas temperatures were then integrated to produce a steady-state three-dimensional model of the disk for which the Stokes parameters were then calculated using the Monte Carlo procedure. In the models presented here, the one-armed oscillation is confined to a region extending out to 10 stellar radii from the central star. We emphasize that the models presented in this section are \textit{ad hoc}: our purpose is to represent plainly the geometric implications of the the density waves such that they can characterize non-axisymmetric disks from a polarimetric perspective.

\begin{figure*}
\epsscale{1.0}
\plotone{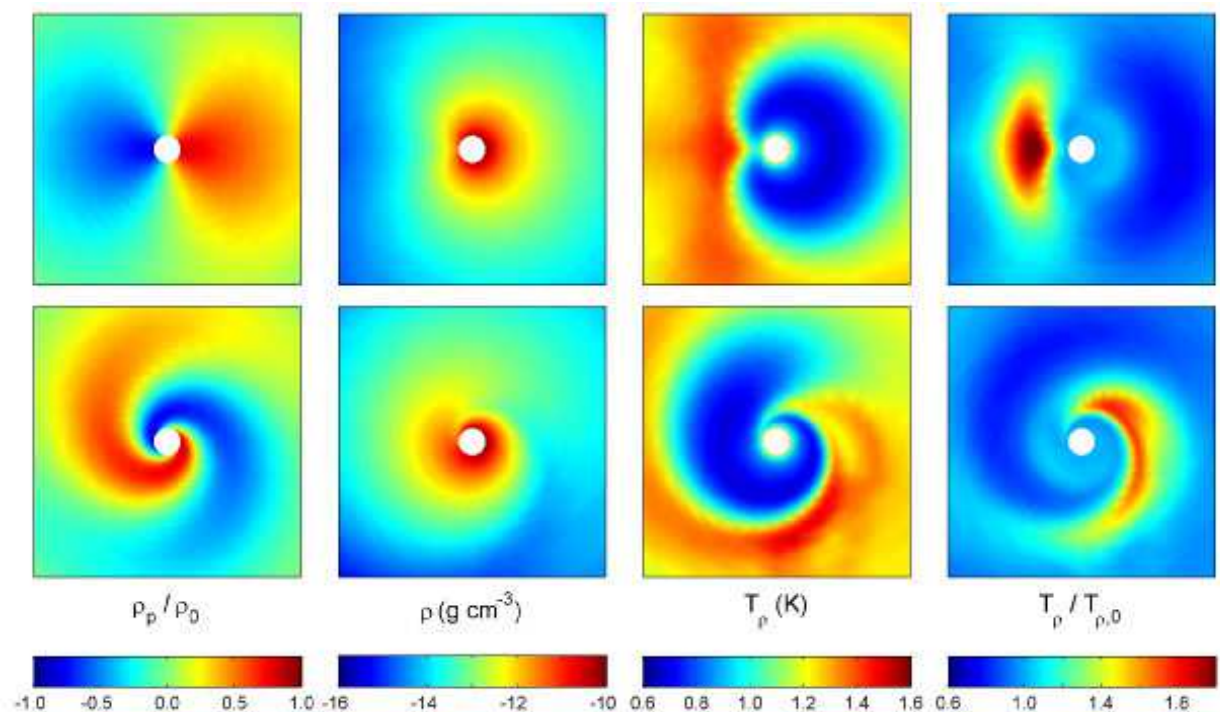}
\caption{Properties of the models of one-armed oscillations in the circumstellar disk surrounding a B2V star. The disk density distribution is parametrized by $n = 3.5$ and $\rho_0 = 5.0 \times 10^{-11}$ g cm$^{-3}$. Each plot represents a slice in the equatorial plane measuring 20 by 20 stellar radii. The top row shows a model characterized by diametrically-opposed overdense and underdense regions (simple oscillation model) while the bottom row shows a model characterized by a logarithmic spiral pattern (spiral oscillation model). First column (leftmost): Density perturbation pattern $\rho/\rho_0$. Second column: Equatorial logarithmic density distribution in the disk (g cm$^{-3}$). Third column: Density-weighted gas temperature ($10^{4}$ K). Fourth column: Ratio of the density-weighted gas temperature of the perturbed disk to that of a disk with an unperturbed density distribution. A color version of this figure is available in the online journal.}
\label{fig:5}
\vspace{0.1in}
\end{figure*}

We present results for the pair of global oscillation models illustrated in Figure \ref{fig:5}. The first perturbation pattern, shown in the top row of Figure \ref{fig:5} and hereafter referred to as the \textit{simple oscillation model}, is characterized by diametrically-opposed overdense and underdense regions similar to the pattern given in \citet{oka97}. The second perturbation pattern, shown in the bottom row of Figure \ref{fig:5} and hereafter referred to as the \textit{spiral oscillation model}, is characterized by a spiral shape similar to the pattern employed in \citet{car09}. The second and third columns of Figure \ref{fig:5} depict the gas density in the equatorial plane and the density-weighted gas temperature, respectively. The last column illustrates the change in gas temperature induced by the presence of the perturbation in the disk relative to an unperturbed model of equal average density. It is important to understand how the inclusion of the perturbation affects the gas density and temperature of the disk. Regions of enhanced or reduced density can affect the conditions for generating the polarization signature in multiple ways. Overdense regions in the disk can increase local opacities and shield parts of the disk to create cool regions of gas where absorption is elevated and scattering is reduced, while underdense regions can decrease local opacities to the point where neither absorption or scattering have an appreciable effect. 

We introduce a phase parameter $\phi_p$ to express the azimuthal angle at which the system is being observed with respect to the perturbation pattern in the disk. We arbitrarily set $\phi_p = 0$ to coincide with the maximum amplitude in the perturbation pattern, and increase $\phi_p$ in the counter-clockwise direction. The minimum amplitude coincides with $\phi_p = \pi$ as the perturbations are antisymmetric. The $\rho/\rho_0$ patterns depicted in Figure \ref{fig:5} are consistent with those presented by \citet{oka97} and \citet{car09} which seem reasonably constrained by observations. The models possess the equivalent mass of an unperturbed disk with density parameters $n = 3.5$ and $\rho_0 = 5.0 \times 10^{-11}$ g cm$^{-3}$. The models discussed were computed with a hydrogen-only composition; we computed the spiral model using a solar composition and found that the temperature structure did not differ appreciably.

\begin{figure*}
\epsscale{1.0}
\plotone{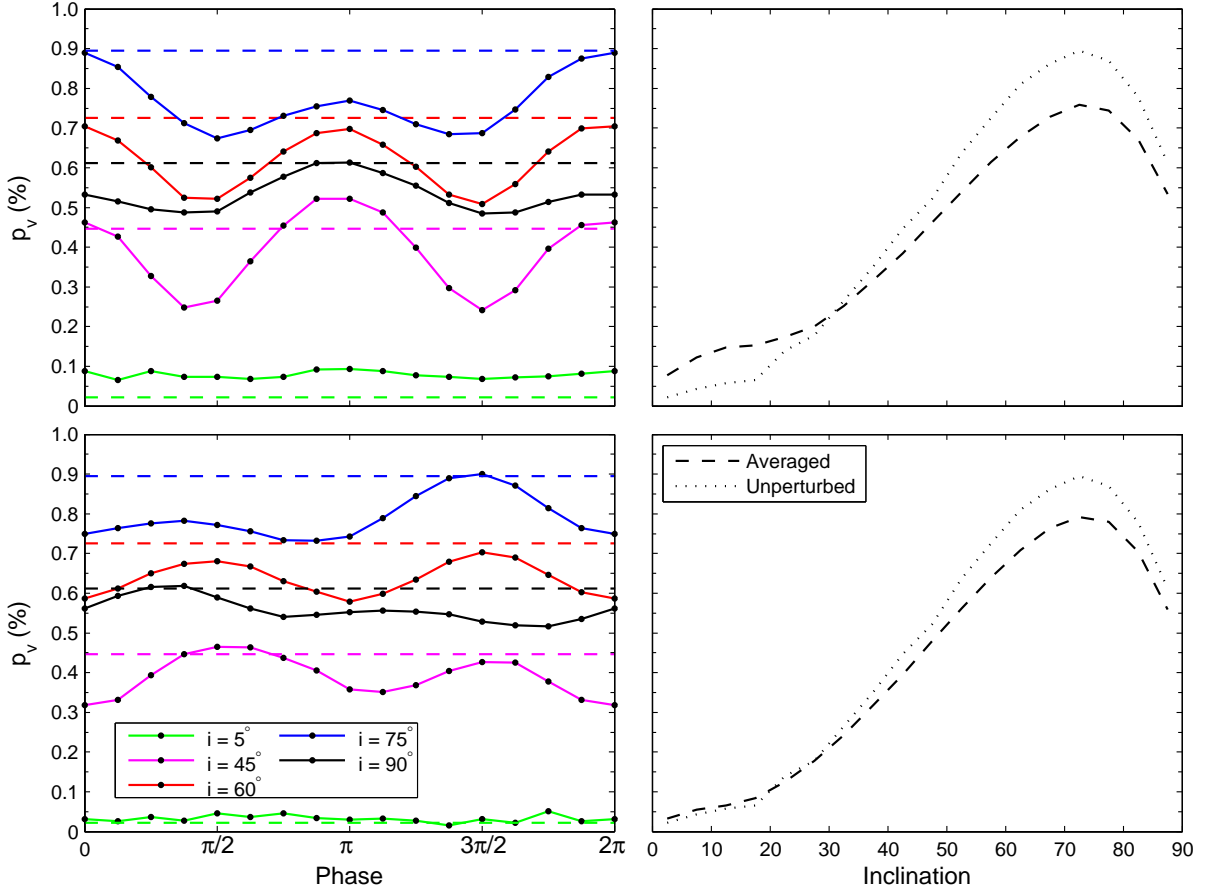}
\caption{Left: Variation in the V-band polarization level with changing phase for the simple oscillation model (top) and the spiral oscillation model (bottom). The solid lines show the fraction of linearly polarized light from the perturbed disk for the system viewed at inclination, plotted from top to bottom,  $75^{\circ}$(blue), $60^{\circ}$(red), $90^{\circ}$(black), $45^{\circ}$(magenta) and $0^{\circ}$(green). The dashed lines show the V-band polarization level from unperturbed disk with the same density distribution. Right: Average V-band polarization level for the perturbed disk and V-band polarization from an unperturbed disk with the same density distribution. A color version of this figure is available in the online journal.}
\label{fig:6}
\vspace{0.1in}
\end{figure*}

Figure \ref{fig:6} shows the V-band polarization levels for the two disk oscillation models and provides a comparison to the polarization level of an unperturbed disk. The average polarization level of the perturbed disk is lower than the polarization level of the unperturbed disk, except for when the disk is viewed at an inclination that is close to pole-on (i.e. $i \lesssim 30^{\circ}$). For near edge-on disks, the variation in the continuum polarization level can exceed 20\% of the average value in the simple oscillation model. In this model, the polarization minima occur at $\phi_p = pi/2$ and $3\pi/2$. Thus, the minimum polarization is produced by the disk when the axis along which the density maximum and minimum are situated is orthogonal to the observer's line-of-sight. As the disk is symmetric about this axis, the polarization level is equivalent at both phases. The polarization maxima occur at $\phi_p = 0$ (or $2\pi$) and $\pi$. The most polarization is produced when the axis along which the density distribution is unperturbed is orthogonal to the observer's line-of-sight. At the phases where this occurs, the observer's line-of-sight can pass through either the overdense region (at $\phi_p = 0$) or the underdense region (at $\phi_p = \pi$). The difference between the polarization level at the two maxima exists because of the combined effects of absorption and stellar occultation of the disk. For the spiral oscillation model, the variation in the continuum polarization level is smaller owing to the more intricate configuration of the underdense and overdense regions. 

\begin{figure}
\epsscale{1.0}
\plotone{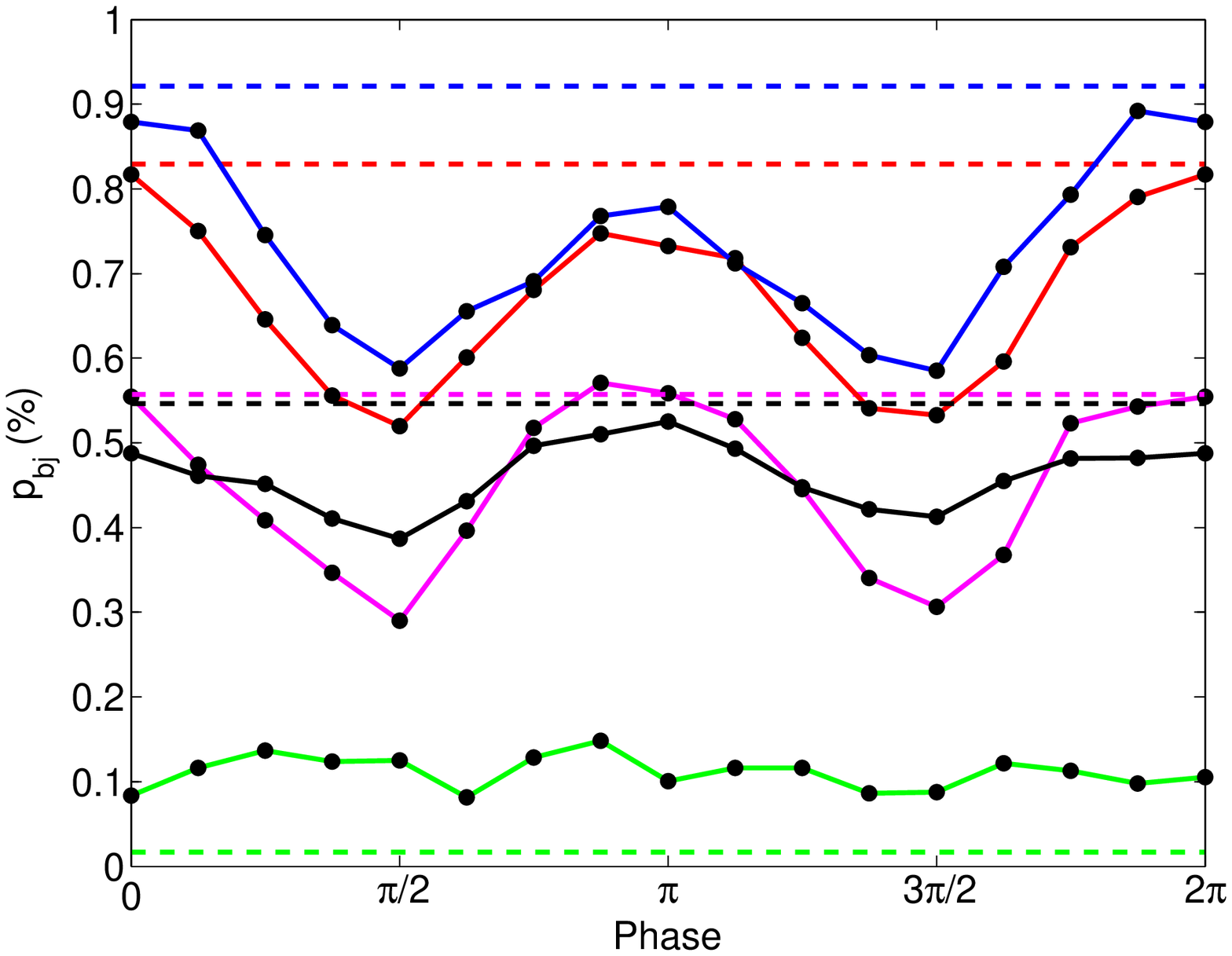}
\caption{Variation in the Balmer jump polarization with changing phase for the simple oscillation model. The solid lines show the fraction of linearly polarized light from the perturbed disk for the system viewed at inclination, plotted from top to bottom,  $75^{\circ}$(blue), $60^{\circ}$(red), $45^{\circ}$(magenta), $90^{\circ}$(black) and $0^{\circ}$(green). The dashed lines show the V-band polarization level from unperturbed disk with the same density distribution. A color version of this figure is available in the online journal.}
\label{fig:7}
\vspace{0.1in}
\end{figure}

\begin{figure}
\epsscale{1.0}
\plotone{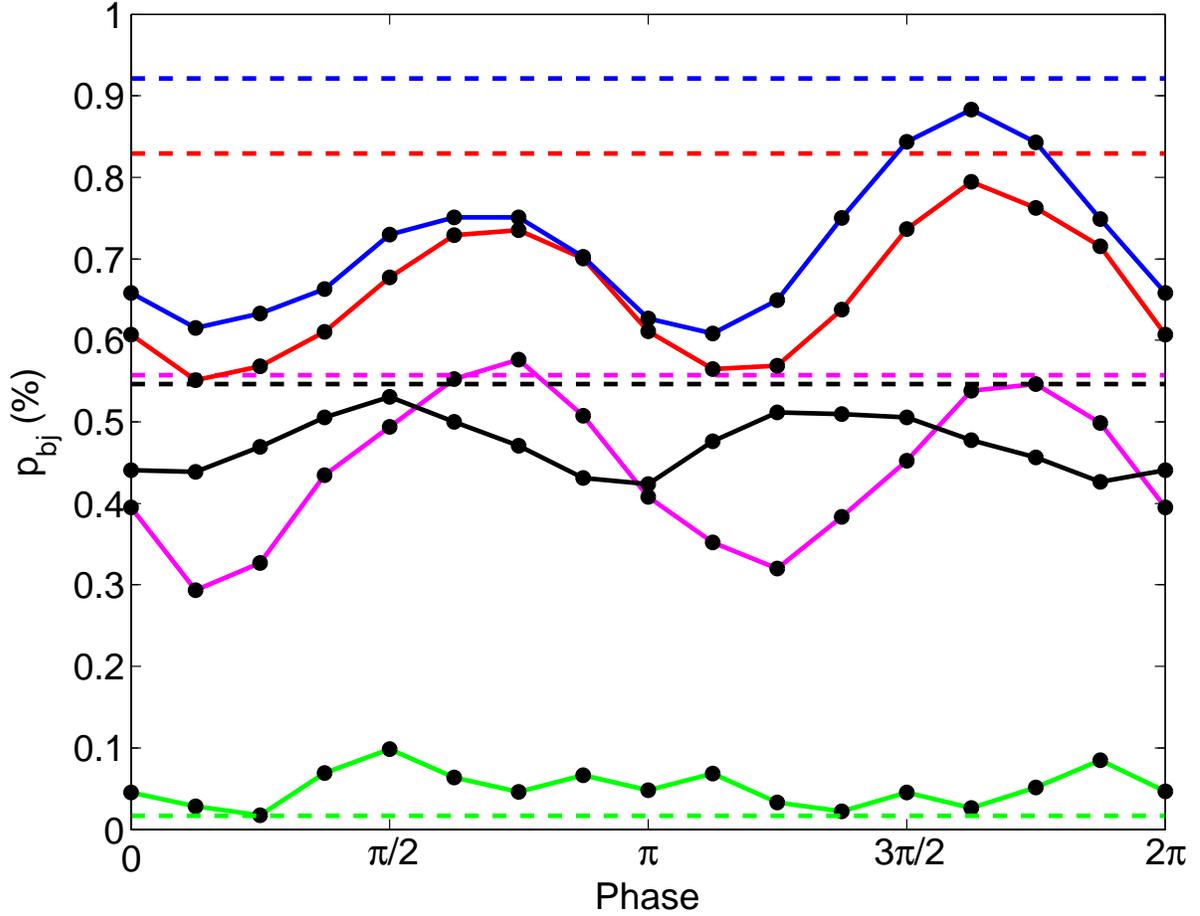}
\caption{Variation in the Balmer jump polarization with changing phase for the spiral oscillation model. The solid lines show the fraction of linearly polarized light from the perturbed disk for the system viewed at inclination, plotted from top to bottom,  $75^{\circ}$(blue), $60^{\circ}$(red), $45^{\circ}$(magenta), $90^{\circ}$(black) and $0^{\circ}$(green). The dashed lines show the V-band polarization level from unperturbed disk with the same density distribution. A color version of this figure is available in the online journal.}
\label{fig:8}
\vspace{0.1in}
\end{figure}

Figures \ref{fig:7} and \ref{fig:8} show the Balmer jumps for the two disk oscillation models. Predictably, the Balmer jumps in the simple oscillation model follow the same variation pattern as the V-band polarization. For the spiral oscillation pattern, the maxima in the Balmer jump are out of phase with the V-band polarization maxima by about $\pi/4$. Of course, this is reflective of the fact that these two polarimetric features arise from different regions of the disk, thus resulting in different azimuthal morphologies at different phases. The Balmer jump arises from a smaller region within 6 $R_*$, while the V-band polarization arises from a larger region of the disk, out to about 10 $R_*$ \citep{car11,hal13b}. Plotting the Balmer jump against the V-band polarization in what is referred to as a BJV diagram \citep{dra11}, the phase differences between the polarimetric features in the two oscillation models are clearly illustrated. While the simple oscillation model sketches straight lines on a BJV plot, the spiral oscillation model traces out discernible loops. We note that for prograde precession of the one-armed oscillation in the disk, the loops will proceed in a clockwise direction in the BJV diagram. 

\begin{figure}
\epsscale{1.0}
\plotone{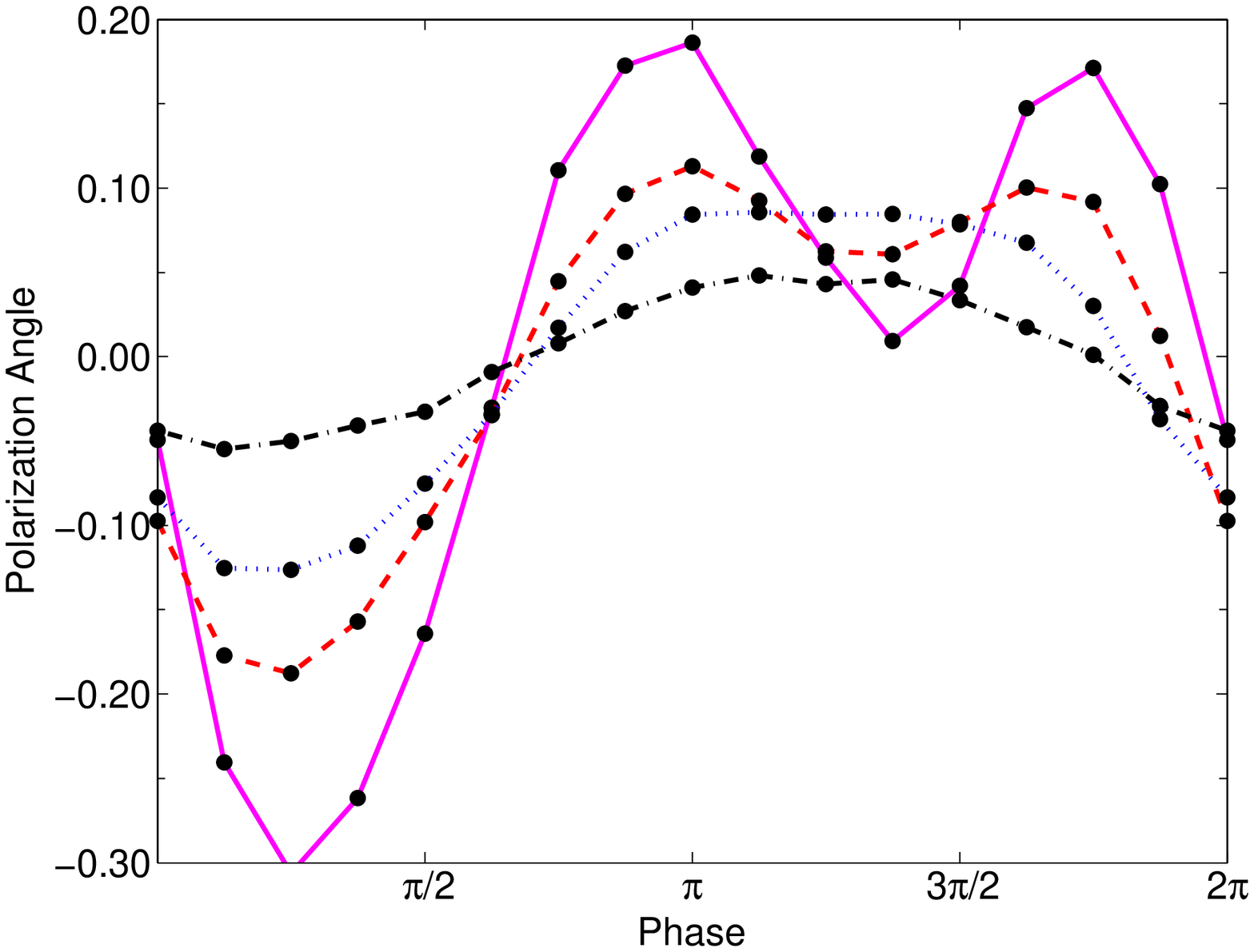}
\caption{Variation in the V-band polarization angle with changing phase for the spiral oscillation model. The lines show the polarization angle of the linearly polarized light from the perturbed disk for the system viewed at inclination $45^{\circ}$(solid, magenta), $60^{\circ}$(dashed, red), $75^{\circ}$(dotted, blue), and $90^{\circ}$(dashed-dotted, black). A color version of this figure is available in the online journal.}
\label{fig:10}
\vspace{0.1in}
\end{figure}

The physical implications of BJV loops in the context of disk growth and dissipation were considered in \citet{hal13b} and \citet{dra11}. When the disk is undergoing such a transition, the BJV loop traces changes to the absorptive opacity and the number of scatterers owing to the addition or removal of gas in the disk. In the case of a precessing, spiral-like density wave, the loop in the BJV diagram appears because of geometric changes to the morphology of the scattering region. Despite the difference in the two scenarios, the principle interpretation of the loop given in \citet{hal13b} is unchanged: a steeper slope in the BJV diagram reflects changes in the absorption imprint whereas a flatter slope reflects changes to the unattenuated polarization level. These changes affect the polarized light that is directed towards the observer as opposed to the total polarized light being produced by the disk, which remains unchanged. Given this explanation, cyclic variations in the polarimetric features owing to one-armed oscillations should provide crucial insight into the structure of the inner disk. We note that while the polarization variation arising from disk growth and dissipation and from a precessing density perturbation are qualitatively similar when plotted in a BJV diagram, the polarization angle varies only for the latter case in which the azimuthal morphology of the scattering medium changes as the perturbation precesses, as illustrated in Figure \ref{fig:10}. The polarization angle remains constant during the formation and dissipation of the disk assuming symmetric addition and removal of gas. Thus, the two phenomenon will be discernible when the polarization is plotted in $q-u$ space, for example. 

\begin{figure}
\epsscale{1.0}
\plotone{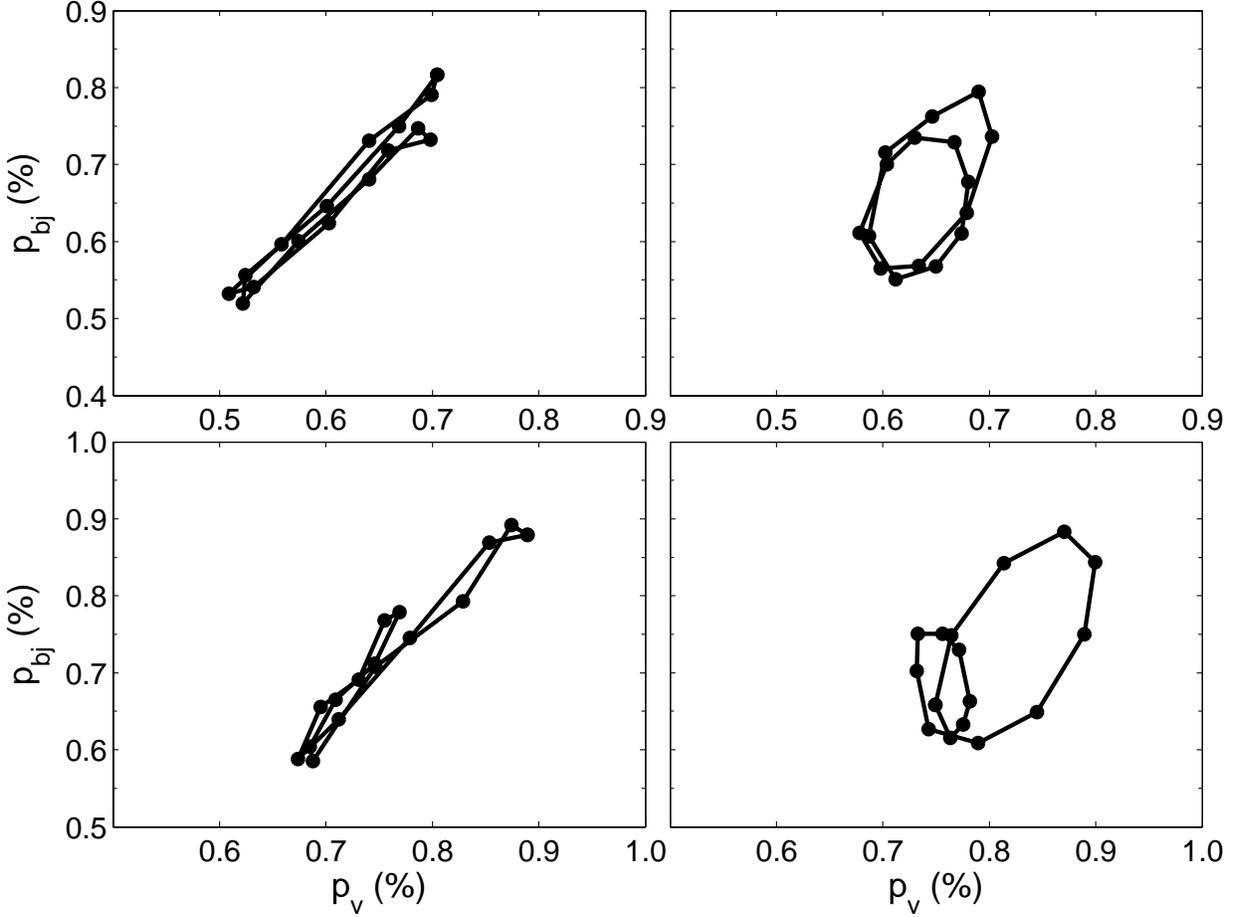}
\caption{Balmer Jump vs. V-band polarization (BJV) diagrams for the disk global oscillation models. The plots on the left show the BJV diagrams for the simple oscillation model while the plots on the right show the BJV diagrams for the spiral oscillation model. The plots on top are for the system viewed at inclination $60^{\circ}$ while those on the bottom are viewed at inclination $75^{\circ}$.}
\label{fig:9}
\vspace{0.1in}
\end{figure}

\section{Summary}

The Balmer jump is an emblematic feature of the polarization signature of classical Be stars and is an important tool for differentiating these objects from similar B-type emission stars. Understanding how this feature behaves is crucial for using it effectively in the study of classical Be stars. We have attempted to ascertain any inherent differences in the polarimetric Balmer jump due to the chemical properties of the system. \citet{wis07b} proposed that an observed lower frequency of polarization Balmer jumps in low-metallicity environments could be explained by higher disk temperatures. Significant differences in the temperature of the circumstellar gas can affect the electron scattering and H{\sc i} absorptive opacities. However, we find that the temperature differences owing to the presence of metals in the disk and to the intrinsic differences in the stellar $T_{\textrm{eff}}$ cannot account for systemically lower Balmer jumps in low-metallicity environments. Wisniewski et al. also suggest that it might be harder to form massive disk systems in low metallicity environments. While we cannot address this point directly, we do point out that classical Be star disks do require sufficient density to produce a detectable wavelength-dependence from the hydrogen absorption signature. Thus, classical Be stars in low-metallicity environments may simply form optically thinner disks than their metal-rich counterparts, thus yielding detectable line emission while exhibiting a linear polarization signature largely unaffected by neutral hydrogen absorption. Clearly, further observations and further study are required to verify and explain the finding of Wisniewski et al.

We also investigated the variability that can occur in the linear polarization signature of classical Be star if the density distribution of the circumstellar gas is not axisymmetric. We have illustrated the qualitative periodic behaviour that we might expect to observe if a one-armed density wave is present in the disk. In particular, polarimetric features originating from different formation regions in the disk could provide key details for mapping the part of the disk close to the star. Few observational programs \citep{mcd00,car09} have investigated possible polarimetric variability in classical Be stars and any correlation it may have to other periodic features whose origins are well-explained by one-armed oscillations. As of yet, no such polarimetric periodicity in classical Be stars has been rigorously demonstrated. It is possible, as Carciofi et al. suggest, that the global oscillation model is a poor predictor of the inner disk geometry, or that mass outbursts that feed the disk disrupt the geometric structure of the inner disk. However, the potential insight that can be gained from variability in the continuum linear polarization signature of classical Be stars suggests further observational study is worthwhile. Good temporal resolution of the polarimetric features paired with reliable hydrodynamical models of one-armed oscillations may be instrumental for determining the geometric nature of the inner region of the circumstellar disks.

\section{Acknowledgements}
CEJ would like to acknowledge support from the Canadian Natural Sciences and Engineering Research Council. The authors would like to thank Jeffrey Bailey for a careful reading of the text as well as the anonymous referee who provided helpful suggestions for improving the manuscript.

\end{document}